\documentclass[preprint,nofootinbib]{revtex4}

\usepackage[centertags]{amsmath}
\usepackage{amsfonts}
\usepackage{amssymb}
\usepackage[dvips]{graphicx}
\usepackage{hyperref}

\DeclareMathOperator{\Sp}{Sp}

\DeclareMathOperator{\Div}{div}

\DeclareMathOperator{\Rot}{rot}

\newcommand{\lan}{\langle}

\newcommand{\ran}{\rangle}

\newcommand{\e}{\varepsilon}

\newcommand{\vf}{\varphi}

\newcommand{\de}{\delta}

\newcommand{\la}{\lambda}

\begin{document}

\title{Relativistic diffusion equation from stochastic quantization}

\date{\today}

\author{P.O. Kazinski}

\email{kpo@phys.tsu.ru}

\affiliation{Department of Physics, Tomsk State University, Tomsk,
634050 Russia}

\begin{abstract}

The new scheme of stochastic quantization is proposed. This
quantization procedure is equivalent to the deformation of an
algebra of observables in the manner of deformation quantization
with an imaginary deformation parameter (the Planck constant). We
apply this method to the models of nonrelativistic and
relativistic particles interacting with an electromagnetic field.
In the first case we establish the equivalence of such a
quantization to the Fokker-Planck equation with a special force.
The application of the proposed quantization procedure to the
model of a relativistic particle results in a relativistic
generalization of the Fokker-Planck equation in the coordinate
space, which in the absence of the electromagnetic field reduces
to the relativistic diffusion (heat) equation. The stationary
probability distribution functions for a stochastically quantized
particle diffusing under a barrier and a particle in the potential
of a harmonic oscillator are derived.

\end{abstract}



\maketitle

\section{Introduction}

There are many different approaches to stochastic quantization and
in understanding what it is (see for the review \cite{Nam}). In
this paper we propose another procedure of stochastic
quantization, which in some sense generalizes the operator
approach to the Fokker-Planck equation used in
\cite{Nam,Par,Ris,ZJ}. This new method of quantization gives a
stochastic mechanics, which is \emph{not equivalent} to quantum
mechanics both in the manner of Nelson's stochastic quantization
\cite{Nel}, and the Parisi-Wu stochastic quantization in the
fictitious time. Rather we interpret the stochastic quantization
from the point of view of the deformation quantization
\cite{BFFLS}, i.e., as a deformation of an associative algebra of
observables (smooth functions over a symplectic manifold) with an
imaginary deformation parameter as opposed to an ordinary quantum
mechanics with a real deformation parameter (the Planck constant).
This formulation of stochastic quantization allows us to apply the
developed methods of quantum mechanics to the stochastic mechanics
almost without any changing.

In this paper we only formulate the general notions of such a
stochastic quantization and show how it works on simple examples:
the models of relativistic and nonrelativistic particles
interacting with an electromagnetic field. The development of the
secondary stochastic quantization and its applications to the
models with infinite degrees of freedom are left for a future
work.

The paper is organized as follows. In the section \ref{stq rules}
we specify the rules of stochastic quantization and introduce
basic notions of the proposed stochastic mechanics. In the section
\ref{examples} we consider two examples: the stochastically
quantized models of a nonrelativistic particle in the subsection
\ref{nonrel part} and a relativistic particle in the subsection
\ref{rel part sec}.

As far as the nonrelativistic case is concerned we find several
simple stationary solutions to the derived equations of motion: a
particle diffuses a potential barrier and a particle in the
potential of a harmonic oscillator. Here we also obtain the
functional integral representation for a transition probability
and the explicit formula for a first correction to the Newton
equations due to the diffusion process. Besides we establish that
the proposed stochastic mechanics can be reproduced by an
appropriate Langevin equation.

In the relativistic case we obtain a Lorentz-invariant
generalization of the Fokker-Planck equation in the coordinate
space, which in the absent of the electromagnetic fields reduces
to the relativistic diffusion (heat) equation (see for the review
\cite{JosPre}). By this example we also show how the basic
concepts of the BRST-quantization (see, e.g., \cite{HeTe}) look in
the context of stochastic mechanics.

In conclusion we sum up the results of the paper and outline the
prospects for possible further research.

\section{The rules of stochastic quantization}\label{stq rules}

In this section we formulate the rules of stochastic quantization
and define the main concepts of such a stochastic mechanics.

Let us given a classical system with the Hamilton function
$H(t,x,p)$, where $x^i$ and $p_j$ are canonically conjugated with
respect to the Poisson bracket positions and momenta
\begin{equation}
    \{x^i,p_j\}=\de^i_j,\qquad i,j=\overline{1,d},
\end{equation}
where $d$ is a dimension of the configuration space. As in quantum
mechanics we associate with such a system the Hilbert space of all
the square-integrable functions depending on $x$ with the standard
inner product
\begin{equation}\label{inner product}
    \langle\psi|\vf\rangle=\int d^dx \psi^*(x)\vf(x),
\end{equation}
Henceforth unless otherwise stated we consider only real-valued
functions in this space.

In the Hilbert space we define the operators $\hat x^i$ and $\hat
p_j$ such that
\begin{equation}\label{comm rel}
    [\hat x^i,\hat p_j]=\hbar\de^i_j,\qquad\hat x^{i+}=\hat
    x^i,\qquad\hat p_j^+=-\hat p_j,
\end{equation}
where $\hbar$ is a small positive number and the cross denotes the
conjugation with respect to the inner product \eqref{inner
product}. Define the Hamiltonian $\hat H(t,\hat x,\hat p)$ by the
von Neumann corresponding rules\footnote{We emphasize that
contrary to \cite{Nam} the Hamiltonian $H(t,x,p)$ is not the
Fokker-Planck Hamiltonian.}
\begin{equation}\label{corresp rules}
    x^i\rightarrow\hat x^i,\qquad p_j\rightarrow\hat p_j.
\end{equation}

The state of the stochastic system is characterized by \emph{two}
vectors $|\psi\ran$ and $|O\ran$ from the Hilbert space with the
evolution
\begin{equation}\label{shrodinger eqs}
    \hbar\frac{d}{dt}|\psi\ran=\hat H|\psi\ran,\qquad\hbar\frac{d}{dt}\lan O|=-\lan O|\hat
    H,
\end{equation}
and the normalization condition
\begin{equation}\label{norm}
    \lan O|\psi\ran=1.
\end{equation}
Define an average of the physical observable $T(t,x,p)$ by the
matrix element
\begin{equation}
    \lan\hat T\ran\equiv\lan O|\hat{T}(t,\hat x,\hat p)|\psi\ran,
\end{equation}
where the operator $\hat{T}(t,\hat x,\hat p)$ is constructed from
$T(t,x,p)$ by the corresponding rules \eqref{corresp rules}. Then
the Heisenberg equations for averages are
\begin{equation}\label{heis eqs}
    \hbar\frac{d}{dt}\lan\hat T\ran=\lan\partial_t\hat T+[\hat T,\hat
    H]\ran.
\end{equation}

By definition the probability density function is
\begin{equation}
    \rho(x)=\lan O|x\ran\lan x|\psi\ran,
\end{equation}
where $|x\ran$ are the eigenvectors for the position operators
corresponding to the eigenvalue $x$. The transition probability
from the position $x$ at the time $t$ to $x'$ at the time $t'$
looks like
\begin{equation}\label{trans prob}
    G(t',x';t,x)=\lan O(t')|x'\ran\lan x'|\hat
    U_{t',t}|x\ran\frac1{\lan O(t)|x\ran},
\end{equation}
where $\hat U_{t',t}$ is the evolution operator obeying the
equations
\begin{equation}
    \hbar\partial_{t'}\hat U_{t',t}=\hat H\hat U_{t',t},\qquad\hat
    U_{t,t}=\hat1.
\end{equation}
The transition probability \eqref{trans prob} possesses the
property of a Markov process
\begin{equation}\label{semigroup prop}
    G(t',x';t,x)=\int d^dy G(t',x';\tau,y)G(\tau,y;t,x).
\end{equation}

By the standard means (see, e.g., \cite{Wein}) we can construct a
path integral representation of the transition probability
\eqref{trans prob}. To this end we introduce auxiliary vectors
$|ip\ran$ in the Hilbert space such that
\begin{equation}\label{p eigenvectors}
    \hat p_j|ip\ran=ip_j|ip\ran,\qquad\lan ip'|ip\ran=\de^d(p-p'),\qquad\int\frac{d^dp}{(2\pi\hbar)^d}|ip\ran\lan ip|=\hat1.
\end{equation}
In the coordinate representation we have
\begin{equation}
    \lan x|ip\ran=\exp{\{-\frac{i}\hbar p_ix^i\}}.
\end{equation}
Then inserting the unity partition \eqref{p eigenvectors} into the
transition probability \eqref{trans prob} we arrive at
\begin{multline}\label{matrix element decomp}
    \lan O(t+dt)|x'\ran\lan x'|\hat U_{t+dt,t}|x\ran\frac1{\lan
    O(t)|x\ran}=\\
    \lan x'|\exp\left\{\frac{dt}\hbar\left[\hat{H}(t,\hat{x},\hat{p}+\hbar\nabla\ln O(t,\hat{x}))+\hbar\partial_t\ln O(t,\hat{x}) \right] \right\}|x\ran=\\
    \int\frac{d^dp(t)}{(2\pi\hbar)^d}\exp{\left\{-\frac{i}\hbar\left[p_i(t)\dot{x}^i(t)+i\left(\bar{H}(t,x(t+dt),ip(t))+\hbar\partial_t
    \ln O(t,x(t))\right)\right]dt\right\}},
\end{multline}
where $x(t)=x$, $x(t+dt)=x'$, $\dot{x}(t)=(x(t+dt)-x(t))/dt$,
$O(t,x)=\lan O(t)|x\ran$ and
\begin{equation}\label{qp symbol}
    \bar{H}(t,x,ip)=\lan x|\hat H(t,\hat p+\hbar\nabla\ln O(t,\hat{x}),\hat x)|ip\ran\lan ip|x\ran
\end{equation}
is a $qp$-symbol of the Hamiltonian $\hat H$ with the momentum
$\hat{p}+\hbar\nabla\ln\hat O$.

The functional integral representation of the transition
probability is obtained by the repeatedly use of the property
\eqref{semigroup prop} and the formula \eqref{matrix element
decomp}:
\begin{multline}\label{trans prob func int}
    G(t',x';t,x)=\int
    \prod_{\tau\in(t,t')}d^dx(\tau)\prod_{\tau\in[t,t')}\frac{d^dp(\tau)}{(2\pi\hbar)^d}\times\\
    \exp\left\{-\frac{i}\hbar\int\limits_t^{t'-d\tau}d\tau\left[p_i(\tau)\dot{x}^i(\tau)+i\left(\bar
    H(\tau,x(\tau+d\tau),ip(\tau))+\hbar\partial_\tau
    \ln O(\tau,x(\tau))\right)\right]\right\}.
\end{multline}
The property \eqref{semigroup prop} guarantees that the functional
integral representation \eqref{trans prob func int} does not
depend on what slices the time interval $[t,t']$ is cut (for more
details see, e.g., \cite{DemCh}).

To conclude this section we formulate the above stochastic
mechanics in terms of the density operator
\begin{equation}\label{density operator}
    \hat\rho=|\psi\ran\lan O|.
\end{equation}
From \eqref{shrodinger eqs} and \eqref{norm} it follows that
\begin{equation}\label{neumann eqs}
    \hbar\frac{d}{dt}\hat\rho=[\hat
    H,\hat\rho],\qquad\Sp\hat\rho=1.
\end{equation}
The averages are calculated as in quantum mechanics
\begin{equation}
    \lan\hat{T}\ran=\Sp(\hat\rho\hat T).
\end{equation}
The probability density function $\rho(t,x)$ is the average of the
projector $|x\ran\lan x|$ and obeys the evolution law
\begin{equation}
    \hbar\partial_t\rho(t,x)=\lan x|[\hat H,\hat\rho]|x\ran.
\end{equation}
As we will see in the next section this equation is nothing but
the Fokker-Planck equation. Notice that from the definition
\eqref{density operator} the density operator is idempotent, i.e.,
\begin{equation}
    \hat\rho^2=\hat\rho.
\end{equation}
By analogy with quantum mechanics one can say that such a density
operator describes a pure state. The transition probability
\eqref{trans prob} is
\begin{equation}
    G(t',x';t,x)=\Sp(\hat\rho(t',t)|x'\ran\lan x'|),\qquad\hat\rho(t,t)=\frac{|x\ran\lan O|}{\lan
    O|x\ran},
\end{equation}
where $\hat\rho(t',t)$ obeys the von Neumann equation
\eqref{neumann eqs}.

The formulation of the stochastic mechanics in terms of the
density operator reveals that from the mathematical point of view
the positions $x^i$ are not distinguished over the momenta $p_j$
as it seems from \eqref{comm rel}. The above stochastic
quantization can be considered as a formal deformation of the
algebra of classical observables in the manner of deformation
quantization \cite{BFFLS}. For a linear symplectic space the Moyal
product is
\begin{equation}
    f(z)*g(z)=\sum\limits_{n=0}^\infty\frac1{n!}\left(\frac{\hbar}2\right)^n\omega^{a_1b_1}\ldots\omega^{a_nb_n}\partial_{a_1\ldots
    a_n}f(z)\partial_{b_1\ldots b_n}g(z),
\end{equation}
where $z\equiv(x,p)$, $a_n,b_n=\overline{1,2d}$, $f(z)$ and $g(z)$
are the Weil symbols, and $\omega^{ab}$ is the inverse to the
symplectic $2$-form $\omega_{ab}$. The trace formula for averages
is given by
\begin{equation}
    \lan\hat{T}\ran=\Sp(\hat{\rho}\hat{T})=\int\frac{d^dxd^dp}{(2\pi\hbar)^d}\sqrt{\det{\omega_{ab}}}\rho(x,p)T(p,x),
\end{equation}
where $\rho(x,p)$ and $T(p,x)$ are $qp$- and $pq$-symbols of the
corresponding operators. For instance, the $qp$-symbol of the
density operator is
\begin{equation}
    \rho(x,ip)=\lan x|\hat{\rho}|ip\ran\lan ip|x\ran.
\end{equation}
Thus all the general results regarding deformation quantization of
symplectic \cite{Fed} and Poisson \cite{Kont} manifolds,
quantization of systems with constraints (see, e.g., \cite{HeTe})
etc. are valid in such a stochastic mechanics.

\section{Examples}\label{examples}
\subsection{Nonrelativistic particle}\label{nonrel part}

In this subsection we consider the stochastic quantization of the
model of a nonrelativistic particle and in particular establish
the one-to-one correspondence of such a quantized model with
appropriate Langevin and Fokker-Planck equations.

According to the general rules expounded in the previous section
the Hamiltonian for a nonrelativistic particle looks
like\footnote{We use the Minkowski metric
$\eta_{\mu\nu}=diag(-1,1,1,1)$ and the system of units in which
the velocity of light $c=1$. The bold face is used for the spacial
components of $4$-vectors.}
\begin{equation}\label{hamiltonian}
    \hat{H}=\frac{(\hat{\mathbf{p}}-\hat{\mathbf{A}})^2}{2m}+\hat{A}^0,
\end{equation}
and the evolution equations \eqref{shrodinger eqs} in the
coordinate representation are
\begin{equation}\label{shrodinger eqs nonrel}
    \hbar\partial_t\psi(t,x)=\left[\frac{(\hat{\mathbf{p}}-\mathbf{A})^2}{2m}-A_0\right]\psi(t,x),\qquad\hbar\partial_tO(t,x)=-\left[\frac{(\hat{\mathbf{p}}+\mathbf{A})^2}{2m}-A_0\right]O(t,x),
\end{equation}
where $\hat{\mathbf{p}}=-\hbar\nabla$ and $A_\mu(t,x)$ are gauge
fields, which we will call the electromagnetic fields. The
physical meaning of the fields $A_\mu$ will be elucidated by the
Fokker-Planck equation associated with \eqref{shrodinger eqs
nonrel}.

The equations \eqref{shrodinger eqs nonrel} are invariant under
the following gauge transformations
\begin{equation}\label{gauge trans}
    \psi(t,x)\rightarrow\psi(t,x) e^{-\vf(t,x)},\qquad O(t,x)\rightarrow O(t,x) e^{\vf(t,x)},\qquad A_\mu(t,x)\rightarrow
    A_\mu(t,x)+\partial_\mu\vf(t,x).
\end{equation}
In particular, these transformations do not change the probability
density function. The conserved $4$-current corresponding to the
gauge transformations \eqref{gauge trans} is
\begin{equation}
    j^\mu=\left(O\psi,\frac1{2m}\left[O(\hat{\mathbf{p}}-\mathbf{A})\psi-\psi(\hat{\mathbf{p}}+\mathbf{A})O\right]\right).
\end{equation}
The system \eqref{shrodinger eqs nonrel} is Lagrangian with the
Hamiltonian action of the form
\begin{equation}
    S_H[O,\psi]=\int
    dtd^dx\left\{\hbar
    O\partial_t\psi-O\left[\frac{(\hat{\mathbf{p}}-\mathbf{A})^2}{2m}-A_0\right]\psi\right\},
\end{equation}
that is the fields $\psi(t,x)$ and $O(t,x)$ are canonically
conjugated.

With the identification
\begin{equation}\label{substitution S}
    O(t,x)\equiv e^{\frac1\hbar S(t,x)},\qquad\psi(t,x)\equiv\rho(t,x)e^{-\frac1\hbar
    S(t,x)},
\end{equation}
the system of evolutionary equations \eqref{shrodinger eqs nonrel}
becomes\footnote{For possible nonlinear generalizations see, e.g.,
\cite{Scar}.}
\begin{equation}\label{FP and qHJ eqs}
    \partial_t\rho=-\Div\left[-\frac{\hbar}{2m}\nabla\rho+\frac{\nabla S-\mathbf{A}}m\rho\right],\qquad\partial_tS-A_0+\frac{(\nabla
    S-\mathbf{A})^2}{2m}=-\frac{\hbar}{2m}\Div(\nabla
    S-\mathbf{A}).
\end{equation}
The first equation in this system is the Fokker-Planck equation,
while the second equation can be referred to as the quantum
Hamilton-Jacobi equation \cite{LMSh}.

Now it is evident that if one neglects quantum corrections then
the initially $\de$-shaped probability density function
$\rho(t,x)$ keeps its own form and propagates as a classical
charged particle in the electromagnetic fields\footnote{Such an
interpretation for the Langevin equation with a non-conservative
force was proposed in \cite{LepMa}.} with particle's momentum
$\nabla S(t,x)-\mathbf{A}(t,x)$.

Let us find the evolution of the average position of the
stochastically quantized particle. The Heisenberg equations
\eqref{heis eqs} for this model are
\begin{equation}\label{Newton eqs}
    m\frac{d}{dt}\lan
    \mathbf{x}\ran=\lan\hat{\mathbf{p}}-\mathbf{A}\ran=\lan\nabla
    S-\mathbf{A}\ran,\qquad
    m\frac{d^2}{dt^2}\lan\mathbf{x}\ran=\lan
    \mathbf{E}\ran+\frac1{m}\lan(\nabla
    S-\mathbf{A})\times\mathbf{H}\ran+\frac\hbar{2m}\lan\Rot{\mathbf{H}}\ran.
\end{equation}
In the case that $\rho(t,x)$ is sufficiently localized comparing
to the characteristic scale of variations of the electromagnetic
fields the angle brackets can be carried through the
electromagnetic fields to obtain a closed system of evolutionary
equations on the average position. They are simply the Newton
equations with the ``quantum'' correction.

Notice that the analog of the quantum mechanical uncertainty
relation is
\begin{equation}
    \lan(x^i)^2\ran\lan(p^{i}_{os})^2\ran\geq\frac{\hbar^2}4,
\end{equation}
where $\mathbf{p}_{os}=-\hbar\nabla\ln\rho^{1/2}$ is the osmotic
momentum. It is easily proven from the inequality
\begin{equation}
    \int d^dx\left[(\xi
    x^i-\hbar\partial_i)\rho^{1/2}\right]^2\geq0,\quad\forall\,\xi\in\mathbb{R}.
\end{equation}
The equipartition law \cite{UlhOrn} can be discovered from
\begin{equation}\label{equipart law}
    \lim_{dt\rightarrow0}T\{\frac{m\dot{\hat{\mathbf{x}}}^2(t)}2dt\}=\frac{m}{2\hbar}[\hat{\mathbf{x}},[\hat{\mathbf{x}},\hat{H}]]=\frac{\hbar}2d,
\end{equation}
where $\hat{\mathbf{x}}(t)$ are the position operators in the
Heisenberg representation and $T$ means the chronological
ordering.

To reproduce the Fokker-Planck equation associated with the
Langevin equation of the form (see, e.g., \cite{ZJ})
\begin{equation}
    \frac{d}{dt}x^i(t)=f^i(t,x(t))+\nu^i(t),\qquad\lan\nu^i(t)\ran=0,\qquad\lan\nu^i(t)\nu^j(t')\ran=\hbar\de^{ij}\de(t-t'),
\end{equation}
where $\nu^i(t)$ is a Gaussian white noise, one has to solve the
system of equations ($m=1$)
\begin{equation}\label{brownian eqs}
    \nabla S(t,x)-\mathbf{A}(t,x)=\mathbf{f}(t,x),\qquad A_0-\partial_tS=\frac12\left(\mathbf{f}^2+\hbar\Div\mathbf{f}\right),
\end{equation}
with respect to $A_\mu(t,x)$ and $S(t,x)$. Obviously, this system
admits a solution. The arbitrariness in the definition of
$A_\mu(t,x)$ and $S(t,x)$ from the equations \eqref{brownian eqs}
is equivalent to the arbitrariness of a gauge. The converse is
also true, i.e., for any given solution $S(t,x)$ and $A_\mu(t,x)$
of the quantum Hamilton-Jacobi equation \eqref{FP and qHJ eqs} we
can construct the force $\mathbf{f}(t,x)$ in the Langevin equation
by the formula \eqref{brownian eqs}, which gives rise to the same
probability distribution function. The equations \eqref{Newton
eqs} for the average position of the particle in the
representation \eqref{brownian eqs} become
\begin{equation}
    \frac{d}{dt}\lan
    \mathbf{x}\ran=\lan\mathbf{f}\ran,\qquad
    \frac{d^2}{dt^2}\lan\mathbf{x}\ran=\lan
    (\partial_t+(\mathbf{f}\nabla))\mathbf{f}\ran+\frac\hbar2\lan\triangle\mathbf{f}\ran.
\end{equation}

To gain a better physical insight into the stochastically
quantized model of a nonrelativistic particle we construct the
functional integral representation \eqref{trans prob func int} of
the transition probability. The $qp$-symbol of the operator
appearing in the formula \eqref{qp symbol} is
\begin{equation}
    \bar{H}(t,x,ip)=\frac1{2m}\left[-\mathbf{p}^2+2i\mathbf{p}(\nabla S-\mathbf{A})-\hbar\Div(\nabla
    S-\mathbf{A})
    \right]+A^0.
\end{equation}
Substituting this expression into \eqref{trans prob func int} and
integrating over momenta we arrive at
\begin{multline}\label{trans prob func int nonrel}
    G(t',x';t,x)=\int
    \left(\frac{m}{2\pi\hbar d\tau}\right)^{d/2}\prod_{\tau\in(t,t')}\left(\frac{m}{2\pi\hbar d\tau}\right)^{d/2}d^dx(\tau)\times\\
    \exp\left\{-\frac1\hbar\int\limits_t^{t'-d\tau}d\tau\left[\frac{m}2\dot{\mathbf{x}}^2+(\mathbf{A}-\nabla S)\dot{\mathbf{x}}-(A^0+\partial_\tau S)-\frac\hbar{2m}\Div(\mathbf{A}-\nabla S) \right]\right\},
\end{multline}
where the functions $A_\mu(t,x)$ and $S(t,x)$ obey the quantum
Hamilton-Jacobi equation \eqref{FP and qHJ eqs} and are taken at
the point $(t,x)=(\tau,x(\tau+d\tau))$. Now it is obvious that the
main contribution to the transition probability is made by the
paths approximating a classical trajectory. In the representation
\eqref{brownian eqs} the transition probability \eqref{trans prob
func int nonrel} reduces to the well known result
\begin{multline}
    G(t',x';t,x)=\int
    \frac{1}{(2\pi\hbar d\tau)^{d/2}}\prod_{\tau\in(t,t')}\frac{d^dx(\tau)}{(2\pi\hbar d\tau)^{d/2}}\times\\
    \exp\left\{-\frac1{\hbar}\int\limits_t^{t'-d\tau}d\tau\left[\frac{(\dot{\mathbf{x}}(\tau)-\mathbf{f}(\tau,x(\tau+d\tau)))^2}2+\hbar\Div\mathbf{f}(\tau,x(\tau+d\tau))\right]\right\}.
\end{multline}

Usually the force $\mathbf{f}(t,x)$ is specified so that the
corresponding Fokker-Planck equation admits a Boltzmann's type
stationary solution. As one can see from the equations \eqref{FP
and qHJ eqs} that is the case if $\nabla S$ and $A_\mu$ are of the
order of $\hbar$ or higher, i.e., the momentum and energy of the
particle are small. For example, the Boltzmann distribution
\begin{equation}
    \rho(x)=e^{-U(x)}/Z,
\end{equation}
where $U(x)$ is some time-independent potential function measured
in terms of the temperature, is reproduced by the following
solution to \eqref{FP and qHJ eqs}
\begin{equation}
    S=-\frac{\hbar U}2,\qquad A_0=\frac{\hbar^2}4\left[\frac12(\nabla
    U)^2+\Delta U\right],\qquad\mathbf{A}=0.
\end{equation}
Possibly such ``quantum'' corrections to the electromagnetic
potential naturally arise from the stochastic quantization of the
electromagnetic fields (we leave a verification of this
supposition for future investigations). Nevertheless in a
high-energy limit, while the diffusion results in small
corrections to the dynamics, the gauge fields $A_\mu$ in the
equations \eqref{FP and qHJ eqs} can be interpreted as the
electromagnetic fields. Notice that under this interpretation the
equations \eqref{FP and qHJ eqs} are Galilean invariant as opposed
to the case, when $\nabla S-\mathbf{A}$ is a force.

To conclude this section we give several simple one-dimensional
stationary solutions to the equations \eqref{FP and qHJ eqs}.

The stationary solutions for $A_\mu=0$. The system of equations
\eqref{FP and qHJ eqs} is
\begin{equation}
    \hbar\rho'=2S'\rho,\qquad\hbar S''+S'^2=2mE,
\end{equation}
where $E$ is a constant. The solutions are
\begin{equation}\label{solution A zero}
\begin{aligned}
    E&=\frac{p^2}{2m}>0,&\quad\rho&=c_1e^{-2px/\hbar}+c_2e^{2px/\hbar}+2\sqrt{c_1c_2},\\
    E&=0,&\quad\rho&=(x-c)^2/Z,\quad\text{or}\quad \rho=c,\\
    E&<0,&\quad\rho&=\frac1Z\cos^2\left[\frac{\sqrt{2m|E|}}\hbar(x-c)\right].
\end{aligned}
\end{equation}
In the last case we can take only one hump of the squared cosine
function and then continue the solution by zero on the residual
part of the line.

To obtain solutions with a finite norm describing a diffusion of
particles under a potential barrier we just have to join the
solutions in \eqref{solution A zero}. For a potential barrier of
the form\footnote{For brevity, we hereinafter designate only
nonvanishing parts of a piecewise function. All the below
solutions have a continuous first derivative on a whole real
line.}
\begin{equation}\label{potential bar}
    A^0(x)=V,\quad x\geq0,
\end{equation}
where $V$ is a positive constant, we have
\begin{equation}\label{solution pot bar}
    \rho=\frac1Z\left\{%
\begin{array}{ll}
    e^{2px/\hbar}(1+p^2l_p^2\hbar^{-2})^{-1}, & x<0; \\
    \cos^2\{[x-l_p\arctan(pl_p\hbar^{-1})]/l_p\}, & 0\leq x<l_p[\arctan(pl_p\hbar^{-1})+\pi/2]; \\
\end{array}%
\right.
\end{equation}
where $0\leq p\leq(2mV)^{1/2}$ and the characteristic penetration
depth
\begin{equation}
    l_p=\frac{\hbar}{\sqrt{2mV-p^2}},
\end{equation}
is of the order of the penetration depth of a quantum mechanical
particle (of course, if one considers $\hbar$ as the Planck
constant). For the potential barrier \eqref{potential bar} there
are normalizable stationary solutions distinct from
\eqref{solution pot bar} of the form
\begin{equation}
    \rho=\frac1Z\left\{%
\begin{array}{ll}
    (x+l_0\cot(a/l_0))^2/l_0^2, & x\in[-l_0\cot(a/l_0),0); \\
    \dfrac{\cos^2[(x-a)/l_0]}{\sin^2(a/l_0)}, & x\in[0,a+\pi l_0/2); \\
\end{array}%
\right.\qquad a\in(0,\pi l_0/2).
\end{equation}

For a small potential barrier
\begin{equation}
    A^0(x)=V,\quad -l/2\leq x<l/2,\qquad l<\pi l_0,
\end{equation}
we obtain the following stationary solutions
\begin{equation}
    \rho=\frac1Z\left\{%
\begin{array}{ll}
    e^{2p(x+l/2)/\hbar}, & x<-l/2; \\
    \dfrac{\cos^2(x/l_p)}{\cos^2[l/(2l_p)]}, & x\in[-l/2,l/2); \\
    e^{-2p(x-l/2)/\hbar}, & x\geq l/2; \\
\end{array}%
\right.\quad
    \rho=\frac1Z\left\{%
\begin{array}{ll}
    (x+c)^2/l_0^2, & x\in[-c,-l/2); \\
    \dfrac{\cos^2(x/l_0)}{\sin^2[l/(2l_0)]}, & x\in[-l/2,l/2); \\
    (x-c)^2/l_0^2, & x\in[l/2,c); \\
\end{array}%
\right.
\end{equation}
where $p$ should be determined from the equation
$p=\hbar\tan[l/(2l_p)]/l_p$ having the unique solution and
$c=l_0\cot[l/(2l_0)]+l/2$. Thus for the barrier of this type the
probability to find a particle near the barrier is higher than
remotely from it.

\begin{figure}[t]
\centering%
\includegraphics*[width=7cm]{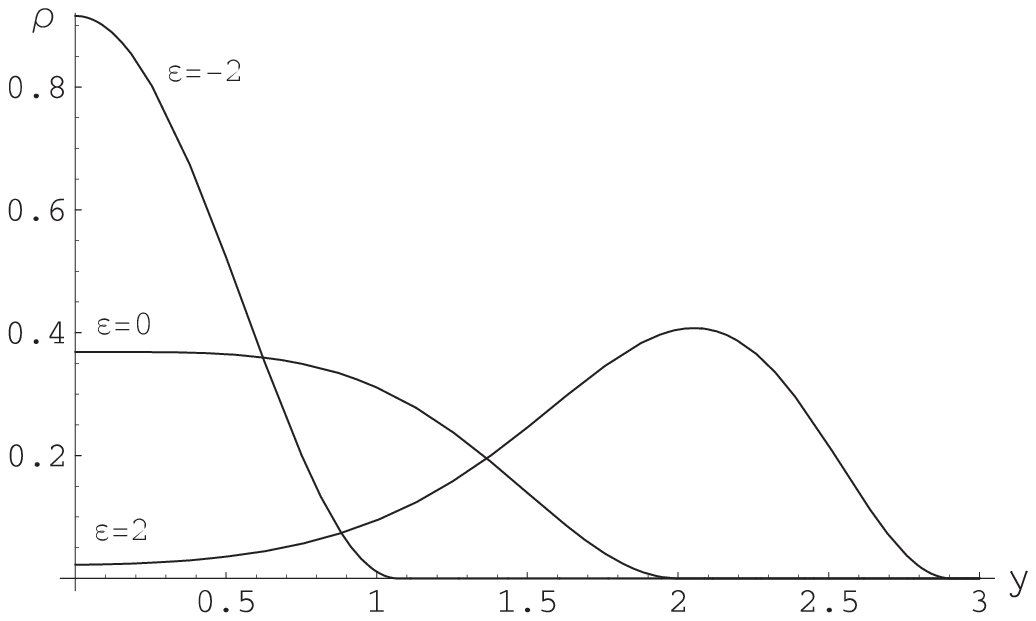}\quad
\includegraphics*[width=7cm]{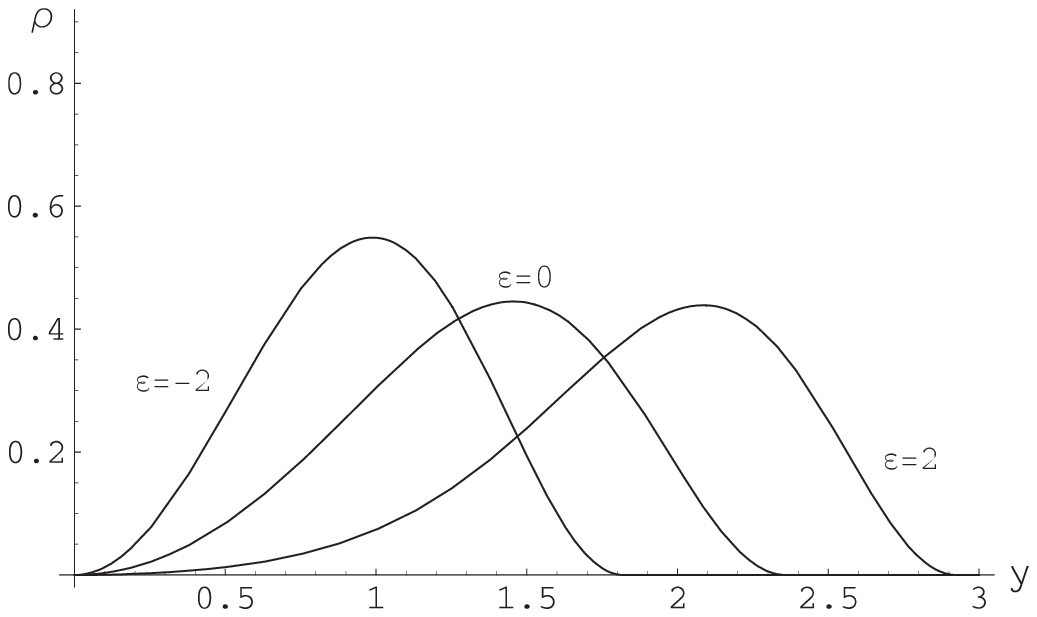}
\caption{{\footnotesize The probability density function for a
stochastically quantized particle in the potential of a harmonic
oscillator. The halves of the first humps normalized on $1/2$ are
only depicted. The solutions corresponding to the first formula in
\eqref{solution harm os} are plotted on the left, while the
solutions corresponding to the second formula in \eqref{solution
harm os} are plotted on the right.}}\label{plots}
\end{figure}

The stationary solutions for $A^0=m\omega^2x^2/2$, $\mathbf{A}=0$.
The system of equations \eqref{FP and qHJ eqs} can be rewritten as
\begin{equation}
    \rho=\frac{O^2}Z,\qquad f''+(y^2-\e)f=0,\qquad
    O(x)=f\left(\left(\frac{m\omega}\hbar\right)^{1/2}x\right),\qquad\e=\frac{2E}{\hbar\omega}.
\end{equation}
Whence from the requirement $\rho'(0)=0$ we have the two types of
stationary solutions
\begin{equation}\label{solution harm os}
    \rho=\frac{e^{-iy^2}}Z\,\Phi^2(\frac{1-i\e}4,\frac12;iy^2),\qquad\rho=\frac{e^{-iy^2}}Zy^2\Phi^2(\frac{3-i\e}4,\frac32;iy^2)\qquad
    y=\left(\frac{m\omega}\hbar\right)^{1/2}x,
\end{equation}
where $\Phi(a,c;x)$ is the confluent hypergeometric function (see,
e.g., \cite{GrRy}). As above we can take only the part of the
solution \eqref{solution harm os} defined on the segment between
two nearest to the minimum of the potential zeros of $\rho(x)$ and
continue it on the residual part of the line by zero. It is
permissible because $\rho(x)$ has degenerate zeroes. Then for an
arbitrary value of the parameter $\e$ these distributions are
bounded and have finite norms (see Fig. \ref{plots}). Otherwise
the integral of $\rho(x)$ diverges logarithmically.

It is not difficult to obtain the asymptotic at
$x\rightarrow+\infty$ of a one-dimensional stationary solution to
\eqref{FP and qHJ eqs} for $A^0=V(x)$, $\mathbf{A}=0$:
\begin{equation}
    \rho\sim\frac{\cos^2[\hbar^{-1}\int(2mV)^{1/2}dx]}{ZV^{1/2}},
\end{equation}
where $V\rightarrow+\infty$ is assumed. The probability density
function $\rho(x)$ has a finite norm if $V(x)$ increases more
rapidly than $x^2$ at both infinities.

\subsection{Relativistic particle}\label{rel part sec}

In this subsection we stochastically quantize the model of a
relativistic particle interacting with the electromagnetic fields.
As the result we obtain a relativistic generalization of the
Fokker-Planck equation in the coordinate space. This model also
serves as a simple example of a model with constraints.

The Hamiltonian action for an interacting relativistic particle
has the form\footnote{In this subsection $d$ is a dimension of the
space-time and $x$ denotes a set of coordinates on it.}
\begin{equation}\label{rel part}
    S_H[x,p,\la]=\int d\tau[p_\mu\dot{x}^\mu-\la((p-A)^2+m^2)],
\end{equation}
where $A_\mu$ is the electromagnetic potential. The dynamics of
the model \eqref{rel part} is governed by the one constraint of
the first kind.

According to the standard BFV-quantization scheme of the models
with constraints of the first kind (see, e.g., \cite{HeTe}) we
introduce a canonically conjugated ghost pair $(c,P)$ and
construct the BRST-charge
\begin{equation}\label{BRST charge class}
    \Omega=c((p-A)^2+m^2),\qquad\{\Omega,\Omega\}=0.
\end{equation}
The quantum BRST-charge is obtained from the classical one by
means of the von Neumann corresponding rules \eqref{corresp
rules}. A graded version of the commutation relations \eqref{comm
rel} for positions and momenta is held. Therefore the quantum
BRST-charge is nilpotent but not Hermitian.

Then the physical state is defined as
\begin{equation}
    \hat\Omega|\psi\ran=0,\qquad\lan
    O|\hat\Omega=0,\;\;\Rightarrow\;\;[\hat\Omega,\hat\rho]=0.
\end{equation}
This definition of a physical state respects also the
BRST-cohomologies structure, i.e., the average over a physical
state of a BRST-exact operator vanishes. Explicitly, in the
coordinate representation we have
\begin{equation}\label{KG eqs}
    [(\hbar\partial_\mu+A_\mu)(\hbar\partial^\mu+A^\mu)+m^2]\psi(x)=0,\qquad[(\hbar\partial_\mu-A_\mu)(\hbar\partial^\mu-A^\mu)+m^2]O(x)=0.
\end{equation}
When the electromagnetic fields vanish these equations are the
Klein-Gordon equations for tachyons\footnote{For the interrelation
between relativistic random walking models and relativistic wave
equations see, for instance, \cite{GJKSch,RanMug}.}.

The action functional for the system of equations \eqref{KG eqs}
is
\begin{equation}
    S[O,\psi]=\int
    d^dxO(x)[(\hbar\partial_\mu+A_\mu)(\hbar\partial^\mu+A^\mu)+m^2]\psi(x).
\end{equation}
As in the nonrelativistic case the action possesses a gauge
invariance under the transformations \eqref{gauge trans}. The
conserved $4$-current looks like
\begin{equation}
    j^\mu=\frac1{2m}\left[O(\hat{p}^\mu-A^\mu)\psi-\psi(\hat{p}^\mu+A^\mu)O\right],
\end{equation}
where $\hat{p}_\mu=-\hbar\partial_\mu$. Making the substitution
\eqref{substitution S} into the system \eqref{KG eqs} we obtain a
Lorentz-invariant generalization of the equations \eqref{FP and
qHJ eqs}
\begin{equation}\label{FP and qHJ rel eqs}
    \partial^\mu\left[-\frac\hbar2\partial_\mu\rho+(\partial_\mu S-A_\mu)\rho\right]=0,\qquad(\partial S-A)^2+m^2=-\hbar\partial^\mu(\partial_\mu S-A_\mu).
\end{equation}
Again the first equation can be called as the relativistic
Fokker-Planck equation in the coordinate space\footnote{For the
relativistic Fokker-Planck equation in the momentum space see,
e.g., \cite{LandLif}. For other approaches to a relativistic
diffusion see, for example, \cite{DunHan,Fa,DTH}.}, while the
second equation is the quantum Hamilton-Jacobi equation. In the
presence of the electromagnetic fields the integral
\begin{equation}\label{norm in rel}
    \int d\mathbf{x}\rho(t,\mathbf{x}),
\end{equation}
is not an integral of motion. Analogously to quantum mechanics we
can explain it by the pair creation.

In the absence of the electromagnetic fields there is a solution
to the quantum Hamilton-Jacobi equation \eqref{FP and qHJ rel eqs}
in the form of a ``plane wave''
\begin{equation}
    S=p_\mu x^\mu,\qquad p^2=-m^2.
\end{equation}
Then the relativistic Fokker-Planck equation is rewritten as
\begin{equation}\label{diffusion rel eqs}
    p^\mu\partial_\mu\rho=\frac\hbar2\Box\rho.
\end{equation}
That is the celebrated relativistic diffusion (heat) equation (see
for the review \cite{JosPre}). It is the hyperbolic type
differential equation and, consequently, the propagation velocity
of small fluctuations does not exceed the velocity of light
contrary to the nonrelativistic diffusion equation. The integral
\eqref{norm in rel} is conserved under an appropriate initial
condition.

Notice that in the same fashion we can quantize the model of a
nonrelativistic particle in the parameterized form
\begin{equation}
    S_H[x,p,\la]=\int d\tau\left[p_\mu\dot{x}^\mu-\la(p_0+H(x,p))\right],
\end{equation}
reproducing the results of the previous subsection.

\section{Concluding remarks}

There are at least two possible points of view on the results of
this paper.

On the one hand we can consider the proposed quantization scheme
from the position of deformation quantization. Then we investigate
in this paper what happens when the algebra of observables is
deformed by an imaginary parameter contrary to quantum mechanics
with the real Planck constant. It would be intriguing if such a
deformation results in a stochastic mechanics related in some way
to real physics. The grounds for these hopes are provided by the
observation that the obtained stochastic mechanics is closely
related to the Langevin and Fokker-Planck equations and in the
classical limit turns into classical mechanics.

On the other hand we can regard the proposed quantization
procedure as another reformulation of the Langevin equation. This
reformulation treats not only nonrelativistic and relativistic
models in a uniform manner, but allows us to extend the developed
methods of quantum mechanics to non-equilibrium statistic physics.

In both cases the work deserves further research. On this way we
can distinguish the secondary stochastic quantization and its
applications to the models with infinite degrees of freedom both
in the relativistic and nonrelativistic cases. The most prominent
models are of course the models of scalar and electromagnetic
fields. Then we can attempt to attack the model of an
incompressible fluid and compare the obtained stochastic model
with the known one for the fully developed turbulence derived from
the Langevin-Navier-Stokes equation (see for the review
\cite{Ant}).

\begin{acknowledgments}

I am grateful to Prof. S.L. Lyakhovich for illuminating
discussions on some aspects of deformation quantization. I
appreciate I.V. Gorbunov and A.M. Pupasov for fruitful debates and
the constructive criticism of the draft of this paper. This work
was supported by the RFBR grant 06-02-17352 and the grant for
Support of Russian Scientific Schools SS-5103.2006.2. The author
appreciates financial support from the Dynasty Foundation and
International Center for Fundamental Physics in Moscow.

\end{acknowledgments}



\begin{thebibliography}{99}


\bibitem{Nam} M. Namiki, \textsl{Stochastic Quantization} (Springer-Verlag, Berlin,
1992); M. Namiki, K. Okano eds., \textsl{Stochastic quantization},
Prog. Theor. Phys. Suppl. No. 111 (1993).

\bibitem{Par} G. Parisi, \textsl{Statistical Field Theory}
(Addison-Wesley, Menlo Park, 1988).

\bibitem{Ris} H. Risken, \textsl{The Fokker-Planck Equation}
(Springer, Berlin, Heidelberg, 1989).

\bibitem{ZJ} J. Zinn-Justin, \textsl{Quantum Field Theory and Critical Phenomena} (Claredon Press, Oxford,
1996).


\bibitem{Nel} E. Nelson, \textsl{Derivation of the Shr\"{o}dinger equation from Newtonian mechanics}, Phys. Rev. \textbf{150}, 1079 (1966); E. Nelson, \textsl{Quantum Fluctuations}
(Princeton University Press, Princeton, New Jersey, 1985).

\bibitem{BFFLS} F. Bayen, M. Flato, C. Fronsdal, A. Lichnerowicz, and D. Sternheimer,
\textsl{Deformation theory and quantization. I. Deformation of
symplectic structures}, Ann. Phys. \textbf{111}, 61 (1978).


\bibitem{JosPre} D.D. Joseph, L. Preziosi, \textsl{Heat waves}, Rev. Mod.
Phys. \textbf{61}, 41 (1989); \textbf{62}, 375 (1990).

\bibitem{HeTe} M. Henneaux, C. Teitelboim, \textsl{Quantization of Gauge Systems} (Princeton University Press, Princeton, New Jersey,
1992).



\bibitem{Wein} S. Weinberg, \textsl{The Quantum Theory of Fields. V.1. Foundations} (Cambridge University
Press, Cambridge, 2000).

\bibitem{DemCh} M. Chaichian, A. Demichev, \textsl{Path Integrals in Physics. V.1. Stochastic Processes and Quantum Mechanics} (Institute of Physics Publishing, Bristol, Philadelphia,
2001).


\bibitem{Fed}  B.V. Fedosov, \textsl{A simple geometrical construction of deformation quantization}, J. Differential Geom. \textbf{40}, 213 (1994).



\bibitem{Kont}  M. Kontsevich, \textsl{Defomation quantization of Poisson manifolds. I}, Lett. Math. Phys. \textbf{66}, 157 (2003). arXiv:q-alg/9709040.

\bibitem{Scar} A.M. Scarfone, \textsl{Stochastic quantization of an interacting classical particle
system}, J. Stat. Mech. 03012, (2007). arXiv:cond-mat/0703115.



\bibitem{LMSh}  G. Litvinov, V. Maslov, and G. Shpiz, \textsl{Idempotent (asymptotic) mathematics and the representation
theory}, arXiv:math/0206025.

\bibitem{UlhOrn} G.E. Uhlenbeck, L.S. Ornstein, \textsl{On the theory of the Brownian motion},
Phys. Rev. \textbf{36}, 823 (1930).

\bibitem{LepMa} D. Leporini, R. Mauri, \textsl{Fluctuations of
non-conservative systems}, J. Stat. Mech. 03002, (2007).


\bibitem{GrRy} I.S. Gradshteyn,
I.M. Ryzhik, \textsl{Table of Integrals, Series and Products. 5th
edition} (Academic Press, Boston, 1994).



\bibitem{GJKSch} B. Gaveau, T. Jacobson, M. Kac, and L.S. Schulman, \textsl{Relativistic extension of the analogy between quantum mechanics
and Brownian motion}, Phys. Rev. Lett. \textbf{53}, 419 (1984).

\bibitem{RanMug} A. Ranfangni, D. Mugnai, \textsl{Stochastic model for tunneling process: The question of superluminal
behavior}, Phys. Rev. E \textbf{52}, 1128 (1995).




\bibitem{LandLif} E.M. Lifshits, L.P. Pitaevskii, \textsl{Physical Kinetics}, (Pergamon, Oxford,
1981).




\bibitem{DunHan} J. Dunkel, P. H\"{a}nggi, \textsl{Theory of relativistic
Brownian motion: The (1+3)-dimensional case}, Phys. Rev. E
\textbf{71}, 016124 (2005).  arXiv:cond-mat/0505532.

\bibitem{Fa} K.S. Fa, \textsl{Analysis of the relativistic Brownian motion in momentum
space}, Braz. J. Phys. \textbf{36}, 777 (2006).



\bibitem{DTH} J. Dunkel, P. Talkner, and P. H\"{a}nggi, \textsl{Relativistic
diffusion processes and random walk models}, Phys. Rev. D
\textbf{75}, 043001 (2007). arXiv:cond-mat/0608023.

\bibitem{Ant} N.V. Antonov, \textsl{Renormalization group, operator product expansion and anomalous scaling
in models of turbulent advection}, J. Phys. A: Math. Gen.
\textbf{39}, 7825 (2006).













\end{thebibliography}
\end{document}